\documentclass[reprint,superscriptaddress,amsmath,amssymb,aps]{revtex4-1}
\usepackage{graphicx}

\usepackage{dcolumn}
\usepackage{upgreek}
\usepackage{gensymb}

\usepackage{txfonts}
\graphicspath{{./}}

\begin{document}

\title{Optical detection of spin orbit torque and current induced heating}

\author{Yukihiro Marui}
\affiliation{Department of Physics, The University of Tokyo, Tokyo 113-0033, Japan}

\author{Masashi Kawaguchi}
\affiliation{Department of Physics, The University of Tokyo, Tokyo 113-0033, Japan}

\author{Masamitsu Hayashi}
\email[]{hayashi@phys.s.u-tokyo.ac.jp}
\affiliation{Department of Physics, The University of Tokyo, Tokyo 113-0033, Japan}
\affiliation{National Institute for Materials Science, Tsukuba 305-0047, Japan}

\newif\iffigure
\figurefalse
\figuretrue

\date{\today}

\begin{abstract}
We have studied spin orbit torque in heavy metal (HM)/ferromagnetic metal (FM) bilayers using magneto-optical Kerr effect. A double modulation technique is developed to separate signals from spin orbit torque and Joule heating. At a current density of $\sim 1 \times 10^{10}$ A/m$^2$, we observe optical signals that scale linearly and quadratically with the current density, both in similar magnitude. The spin orbit torque estimated using this technique is consistent with that evaluated using spin transport measurements. We find that changes in the refractive index of the film with temperature is the main source of the heating induced signal.
\end{abstract}

\pacs{}

\maketitle

Generation of spin current via the spin Hall effect\cite{N2388} in heavy metals (HM) and spin accumulation by the Rashba-Edelstein\cite{N2070} effect at interfaces are attracting great interest owing to their potential application toward magnetic random access memory technologies. The spin current and/or spin accumulation can diffuse into adjacent magnetic layer(s) to exert spin orbit torque\cite{N1895,N1896,N2927} on magnetic moments and allow current induced control of magnetization. Advances in the understanding of the mechanism of how spin orbit torque arise at interfaces has been made owing to measurement techniques developed based on electrical\cite{N1357,N1224,N1902,N2566,N2831} or optical measurements\cite{N2907,N3374}. The accuracy of the torque evaluated using different techniques, however, seems to differ depending on the system studied and its improvement remains as a subject to be addressed\cite{N3371,N3378}.

Recent experiments have demonstrated that it is possible to optically detect spin accumulation at sample edges or surfaces generated by the spin Hall effect in semiconductors\cite{N125} and metals\cite{N3027,N2908,N3363,N3362}. Direct measurements of the spin accumulation allows straightforward characterization of the efficiency of spin current generation in various materials. However, it has been pointed out that the optical signal used to study spin accumulation can be contaminated by current induced heating (Joule heating) effects\cite{N3366,N3361}. To resolve signals originating from current induced spin accumulation, one needs to reduce the current density to minimize Joule heating effects and simultaneously improve the signal to noise ratio of the optical detection setup\cite{N3363}.

Here we show that it is possible to separate magnetic and heating induced signals in optical measurements. In model systems consisting of HM/ferromagnetic metal (FM) bilayer, we study spin orbit torque using a double modulation magneto-optical detection technique. We find signals that scale linearly and quadratically with the current density. The former is due to current induced changes of the magnetization, which reflects the size of spin orbit torque. The estimated torque is in relatively good agreement with transport measurements of the same sample. The signal that scales quadratically with the current density is larger than the signal associated with the spin orbit torque even at a current density of $\sim 1 \times 10^{10}$ A/m$^2$. We find that it is the changes in the refractive index of the film (and/or the silicon oxide layer of the substrate) with temperature that causes the spurious heating induced signal. The detection scheme developed here may be applied to detect current induced spin accumulation at surfaces and interfaces using larger current density, providing easier access to study such effects.

\begin{figure}
	\includegraphics[width=0.49\linewidth]{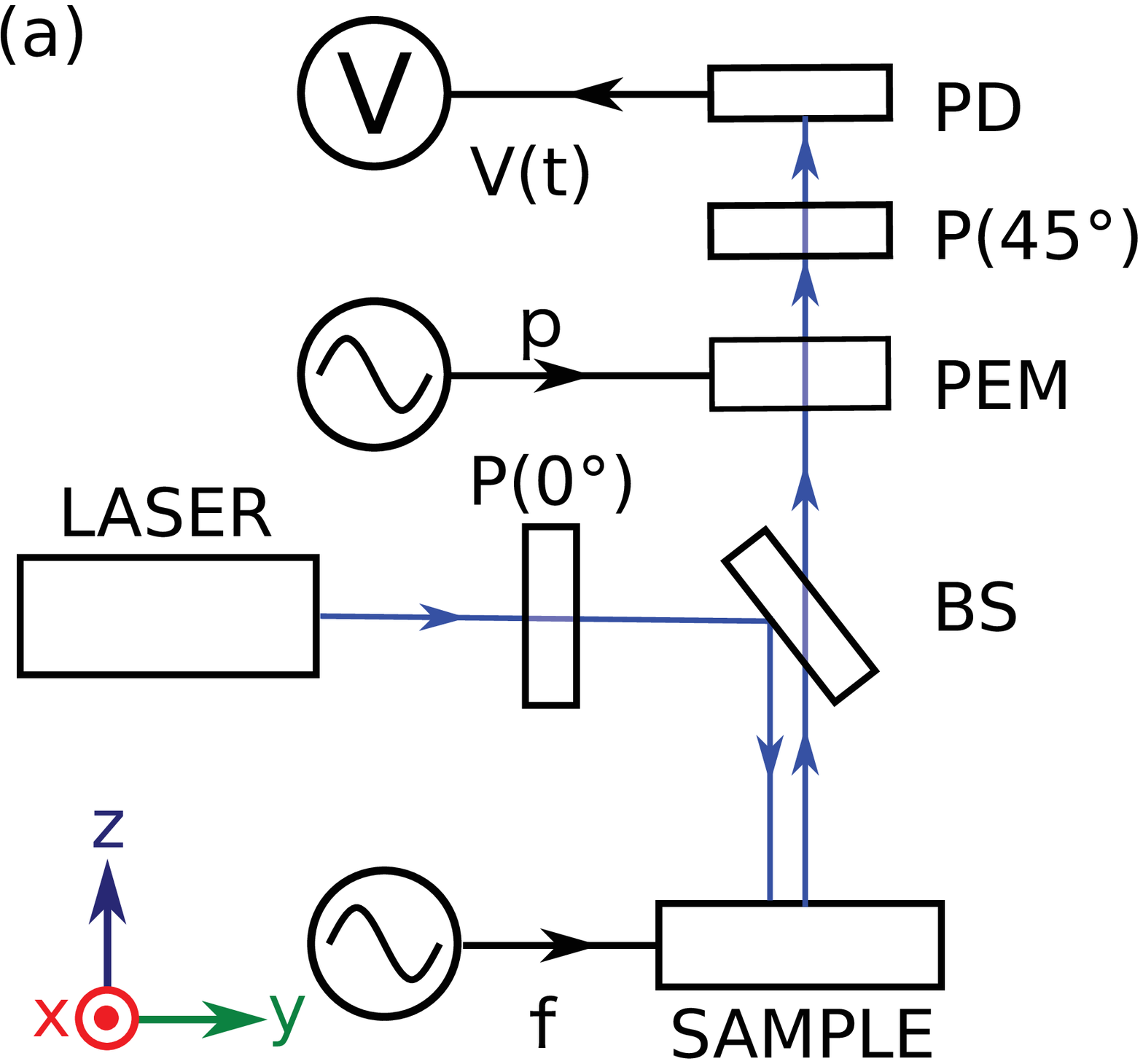}
	\includegraphics[width=0.49\linewidth]{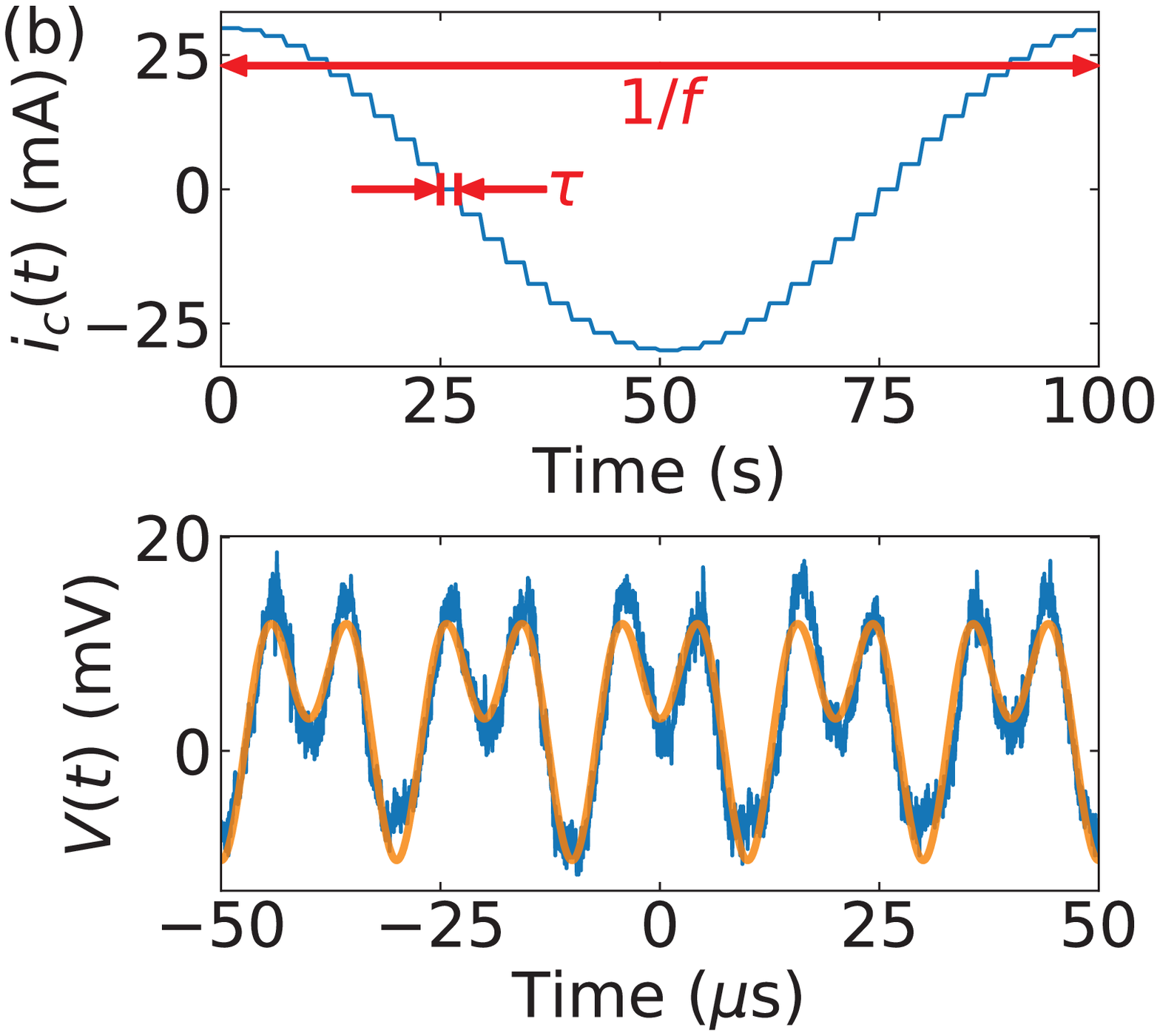}
	\caption{(a) Schematic illustration of the experimental setup and the coordinate axis. P: polarizer, BS: beam splitter, PD: photo detector, PEM: photo elastic modulator. (b) Temporal variation of the current $i_{c}$ passed to the sample (upper panel) and an exemplary time-expanded plot of the voltage $V(t)$ from the photo detector measured using an oscilloscope (bottom panel, blue and orange lines represent experimental results and calculation using Eq. (\ref{eq:oscilloscope}), respectively). Note that the voltage shown here is amplified so that it can be monitored with an oscilloscope.}
	\label{fig:configkerr-setup}
\end{figure}

Samples are grown on thermally oxidized silicon substrates using RF magnetron sputtering. Two representative  films are made. A: sub./0.5 Ta/3 Pt/1 Co$_{20}$Fe$_{60}$B$_{20}$/2 MgO/1 Ta and B: sub./3 W/1 Co$_{20}$Fe$_{60}$B$_{20}$/2 MgO/1 Ta (thickness in nanometer). Films are deposited through a metal shadow mask to create wires and Hall bars. The width and length (i.e. distance between the voltage probes) of the wires and Hall bars are 0.4 mm and 1.2 mm respectively. All films are deposited and measured at room temperature. The magnetization easy axis points along the film plane (we find little perpendicular magnetic anisotropy at the CoFeB/MgO interface without any post annealing).

The experimental setup and the coordinate axis are schematically illustrated in Fig. \ref{fig:configkerr-setup}. The film normal points along the $z$ axis and the current is passed along the $y$ axis. Positive current is defined as current flow to $+y$. Magneto-optical Kerr effect is used to probe the magnetization direction of the films. A continuous wave (CW) He-Ne laser (wavelength: 633 nm, power: 0.5 mW) is used as the light source. The light is polarized along the $x$ axis and is irradiated to the sample (at the center of the wire) from the film normal. The reflected light from the sample goes through a photo elastic modulator (PEM) and a polarization filter before entering a silicon avalanche photo detector. The modulation frequency $p$ of the PEM is $\sim 50 \mbox{ kHz}$.

As the incident light is polarized along the $x$ axis, we express its polarization as $\mathbf{E}_{\textrm{IN}} = E_0 (1\mathbf{e}_{x} + 0 \mathbf{e}_{y})$, with $\mathbf{e}_{i}$ representing an unit vector along the $i$ direction and $E_0$ is the intensity of the light. 
The polarization of the reflected light that enters the silicon photodetector takes the following form:
\begin{align}\label{IncidentReflected}
	\mathbf{E}_{\textrm{PD}} = \hat{\textrm{H}}_{\textrm{P45}} \hat{\textrm{H}}_{\textrm{PEM}} \hat{\textrm{H}}_{\textrm{K}} \mathbf{E}_{\textrm{IN}}
\end{align}
where $\hat{\textrm{H}}_{\textrm{K}}$ represents the magneto-optical Kerr effect of the sample\cite{N3371}, $\hat{\textrm{H}}_{\textrm{P45}}$ and $\hat{\textrm{H}}_{\textrm{PEM}}$ are the operations of the polarization filter and the PEM placed after the light is reflected from the sample. The matrix form of these operations are
\begin{align}\label{HK}
	\hat{\textrm{H}}_{\textrm{K}} =
	r_{K}\begin{bmatrix}
		\cos\theta \cos\eta-i \sin\theta \sin\eta & -\sin\theta \cos\eta-i \cos\theta \sin\eta \\
		\sin\theta \cos\eta + i \cos\theta \sin\eta & \cos\theta \cos\eta - i \sin\theta \sin\eta  
	\end{bmatrix}
\end{align}
\begin{align}\label{HPEM}
	\hat{\textrm{H}}_{\textrm{PEM}} =
	\begin{bmatrix}
		1 & 0                            \\
		0 & e^{-i \delta\sin(2\pi pt)} 
	\end{bmatrix}
\end{align}
\begin{align}\label{HP45}
	\hat{\textrm{H}}_{\textrm{P45}} =
	\frac{1}{2}
	\begin{bmatrix}
		1 & 1 \\
		1 & 1 
	\end{bmatrix}
\end{align}
$ r_K $ represents the reflectivity, $ \theta $ and $ \eta $ are the Kerr rotation angle and the ellipticity of the sample. $\delta$ is a phase delay at the PEM.

The light intensity ($I_{\textrm{PD}}= |{\mathbf{E}_{\textrm{PD}}}|^{2}$) that arrives at the silicon photodetector is
\begin{align}\label{eq:oscilloscope}
	I_{\textrm{PD}} = \frac{(r_KE_{0})^2}{4} [1+\sin(2\theta)\cos(2\eta)\cos(\delta\sin(2\pi  pt)) & \nonumber \\
	+\sin(2\eta)\sin(\delta\sin(2\pi pt))                                                    & ]         
\end{align}
According to Eq. (5), $ I_{PD} $ oscillates with frequency $p$. We Fourier expand $I_{\textrm{PD}}$ in series of the PEM frequency $p$:
\begin{align}\label{FourierExp}
	I_{PD} = \frac{(r_KE_{0})^2}{2} \sum_{n} [I_n \cos(2\pi n pt) + \tilde{I}_n \sin(2\pi n pt)]
\end{align}
The first three components ($n=0,1,2$) are expressed as
\begin{alignat}{4}
	 & I_0  = \frac{1}{2}(1+J_{0}(\delta)\sin(2\theta)\cos(2\eta)), \tilde{I}_0=0\\
	 & I_1 = 0,\ \tilde{I}_1 = J_{1}(\delta)\sin(2\eta)              \\
	 & I_2 = J_{2}(\delta)\sin(2\theta)\cos(2\eta),\ \tilde{I}_2=0 
\end{alignat}
$J_n(x)$ is the Bessel function of the first kind. The PEM phase delay $\delta$ is set to $\sim 2.4 \mbox{ rad}$ to obtain $ J_{0}(\delta) \sim 0 $, which results in $I_0  \sim \frac{1}{2}$. With $\theta \ll 1$ and $\eta \ll 1$, we obtain $\tilde{I}_1 \sim 2 \eta J_1(\delta)$ and $I_2 \sim 2 \theta J_2(\delta)$. 

The light is converted to electrical voltage at the photo detector: the conversion process is linear. The output voltage from the photo detector is fed into a lock-in amplifier to pick up the two frequency components $\tilde{I}_{1}$ and $I_{2}$. The DC component $I_0$ is measured with a digital multimeter. A representative trace of the voltage that enters the lock-in amplifier, which shows the modulation by PEM, is shown in Fig. \ref{fig:configkerr-setup}(b), lower panel.
From the Fourier transform of the data, the Kerr rotation angle and the ellipticity are obtained by the following identities:
\begin{align}\label{Rotation}
	\theta \sim \frac{1}{4 J_{2}(\delta)}\frac{I_2}{I_0}
\end{align}
\begin{align}\label{Ellipticity}
	\eta \sim \frac{1}{4 J_{1}(\delta)}\frac{\tilde{I}_1}{I_0}
\end{align}
with $J_{1}(\delta) \sim 0.52$ and $J_{2}(\delta) \sim 0.43$ for $\delta \sim 2.4 \mbox{ rad}$. $\theta$ and $\eta$ reflect the magnetic state of the sample. Since both quantities $\theta$ and $\eta$ provide the same information, from hereafter we focus on $\eta$ which is larger in magnitude than $\theta$ here (the relative size of $\theta$ and $\eta$ is mostly determined by the light wavelength and the thickness of the silicon oxide layer of the substrate\cite{N3337}). 

In general, spin orbit torque can be decomposed into two components, i.e. the damping-like and the field-like components\cite{N1902,N2566,N2408,N2927}. When current is passed along $+y$ and the initial magnetization is directed along $+y$, the effective field associated with the damping-like $h_{DL}$ and field-like $h_{FL}$ components of the torque points along $\pm z$ and $\pm x$, respectively. Here we study the damping-like component which is known to predominantly originate from the spin transfer torque\cite{N280} when spin current is injected into the FM layer via the spin Hall effect of the HM layer. We thus use the polar geometry, as shown in Fig. \ref{fig:configkerr-setup}, to probe the out of plane component $m_z$ of the magnetization in response to current along $y$.

\begin{figure}
	\includegraphics[width=0.375\linewidth]{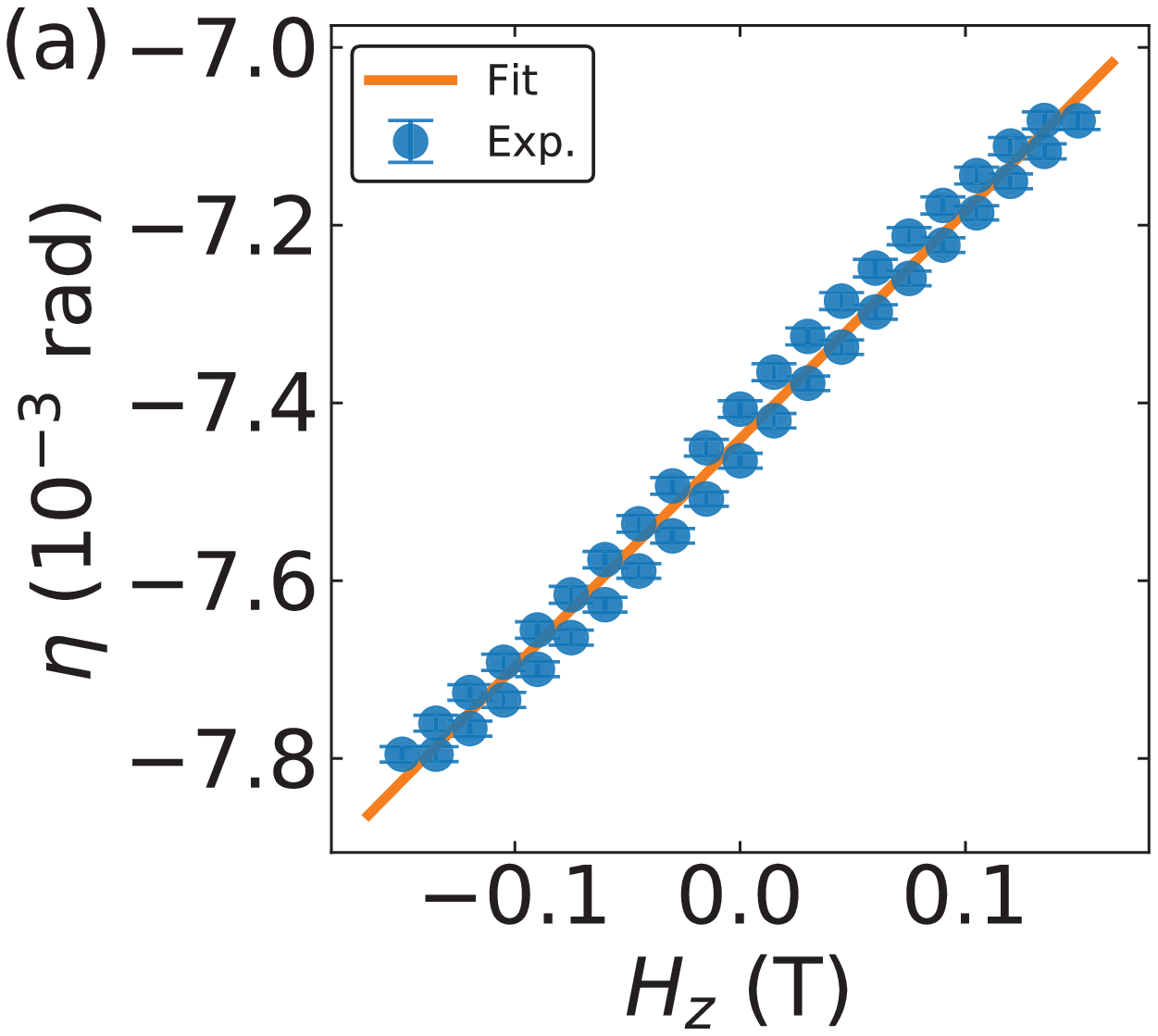}
	\includegraphics[width=0.6\linewidth]{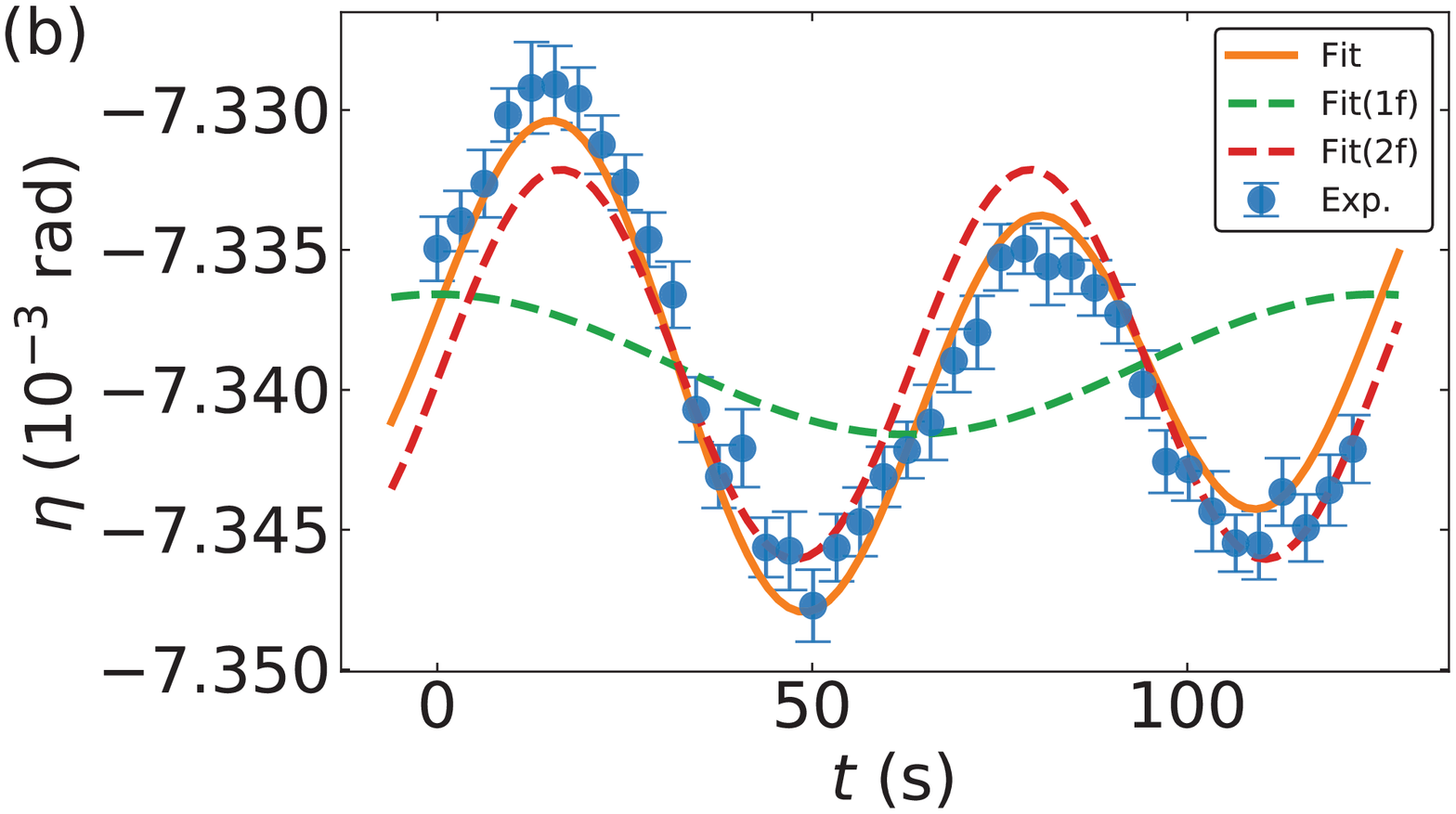}
	\caption{(a) $H_z$ dependence of the Kerr ellipticity ($\eta$). The red solid line shows a linear fit to the data. (b) Temporal variation of $\eta$ when $i_{c}$ is passed. $i_{c}$ is modulated with frequency $f$. Here $i_{0} = 16$ mA and $f = 0.8\times10^{-2}$ Hz ($ \tau =3$ s). The solid line shows fit to the data with Eq. (\ref{EtaT}). The dashed lines show contributions from $\eta_{1f}$ and $\eta_{2f}$ terms. The error bars reflect the standard deviation of the measurements carried out repeatedly. Results are from sample B.}
	\label{fig:wave}
\end{figure}

To convert the optical signal ($\eta$) to an effective field, we first study the magnetic field response of the magnetization as a function of an out of plane field $H_z$. Using the relation noted in Eq. (\ref{Ellipticity}), the Kerr ellipticity is measured as a function of $H_z$: the results are plotted in Fig. \ref{fig:wave}(a). The change of $\eta$ with $H_z$ is almost linear. We fit the data with a linear function to obtain a proportionality constant $\alpha$ which can be used to convert $\eta$ to an effective field (see Eq. (\ref{alpha}) shown below). $\alpha$ is listed in Table \ref{tab:efficiency} for both samples (A and B). Note that the offset in $\eta$ is predominantly due to an intentional tilting of $\mathbf{E}_{\textrm{IN}}$ polarization with the $x$ axis (i.e. the PEM polarization plane). Without the tilting, the optical signal becomes distorted and hinders accurate measurements of $\eta$.

Next we study the effect of current on the magnetization. We apply a constant current $i_{c}$ to the sample for a certain duration $\tau$ to acquire $\eta$. The size of $i_{c}$ is varied stepwise to follow a sinusoidal function with frequency $f$ (see Fig. \ref{fig:configkerr-setup}(b), upper panel), i.e. $i_{c}(t)=i_{0} \cos(2\pi f t)$. Since $\eta$ is evaluated at each $i_{c}$, we denote $\eta$ as $\eta(t)$ hereafter to explicitly show its time dependence through $i_{c}(t)$. This process of varying $i_{c}$ with frequency $f$, i.e. the second modulation (the first one is at the PEM), is repeated many times (typically 1000 times) to improve the signal to noise ratio. The experiment was conducted for various values of $i_{0}$. We compute the equivalent current density $j$ that passes through the HM layer with $i_{0}$. To calculate $j$, the resistivity of each layer is assumed as the following. Ta: $\sim$ 200 $\mu \Omega$ cm, $W: \sim$ 110 $\mu \Omega$ cm, Pt: $\sim$ 55 $\mu \Omega$ cm, and CoFeB: $\sim$ 160 $\mu \Omega$ cm (Ref. \cite{N2657}). During the measurements with current, a magnetic field directed along the $y$ axis is applied ($H_y \sim \pm$ 50 mT) to avoid causing demagnetization and also to set the initial magnetic state of the wire. 

A representative data of $\eta (t)$ is shown in Fig. \ref{fig:wave}(b). The time dependent Kerr ellipticity $\eta (t)$ is fitted with a sum of sinusoidal functions with frequency $f$ and $2f$, i.e.
\begin{align}\label{EtaT}
	\eta(t) = \eta_{1f} \cos(2\pi ft + \varphi_{1f}) + \eta_{2f} \cos(4\pi ft+\varphi_{2f}) + \eta_{0}
\end{align}
In Eq. (\ref{EtaT}), $\eta_{1f}$ is the term that scales linearly with current and represents contribution from the spin orbit torque whereas $\eta_{2f}$ is proportional to the square of current and is likely related to Joule heating induced effects. $ \eta_{0} $ is the background signal described above.
$ \varphi_{1f} $ and $ \varphi_{2f}$ are the phase delay that will be discussed later. Contribution from each component ($\eta_{1f}$ and $\eta_{2f}$) is shown in the figure by the dashed lines. As evident, the $2f$ component that scales with the square of $j$ dominates the signal.

Using $\alpha$ listed in Table \ref{tab:efficiency}, we convert $\eta_{1f}$ and $\eta_{2f}$ to effective fields using the following relation:
\begin{align}\label{alpha}
	\begin{split}
		\eta_{1f} = \alpha h_{DL} \\
		\eta_{2f} = \alpha h_{HE}
	\end{split}
\end{align}
where $h_{HE}$ represents the size of Joule heating induced effects. Note that $h_{HE}$ is not a field that appears due to Joule heating; it is to characterize the effect so that we may compare its influence on the optical signal with that of current induced torques.

\begin{figure}[t]
	\includegraphics[width=0.49\linewidth]{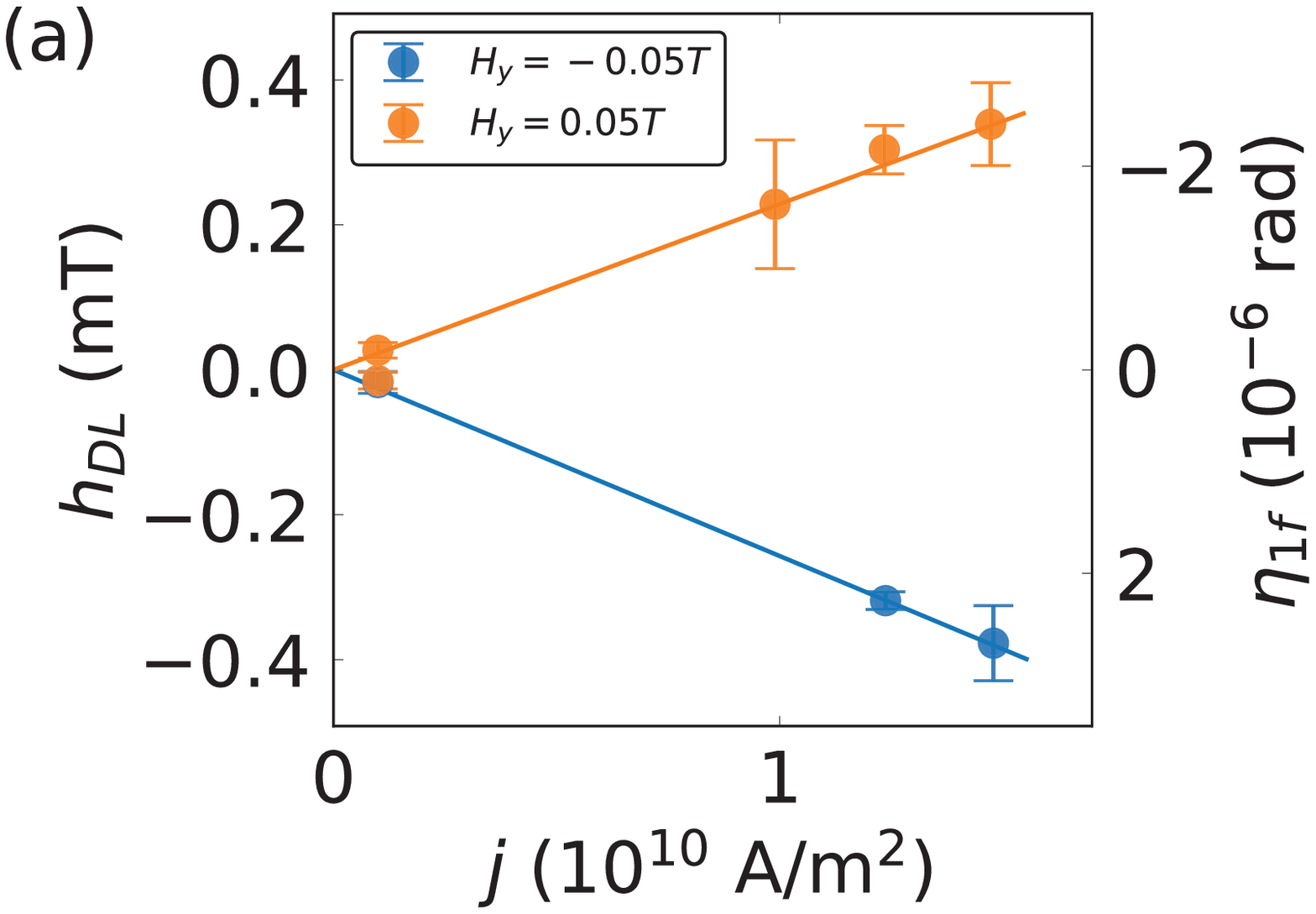}
	\includegraphics[width=0.49\linewidth]{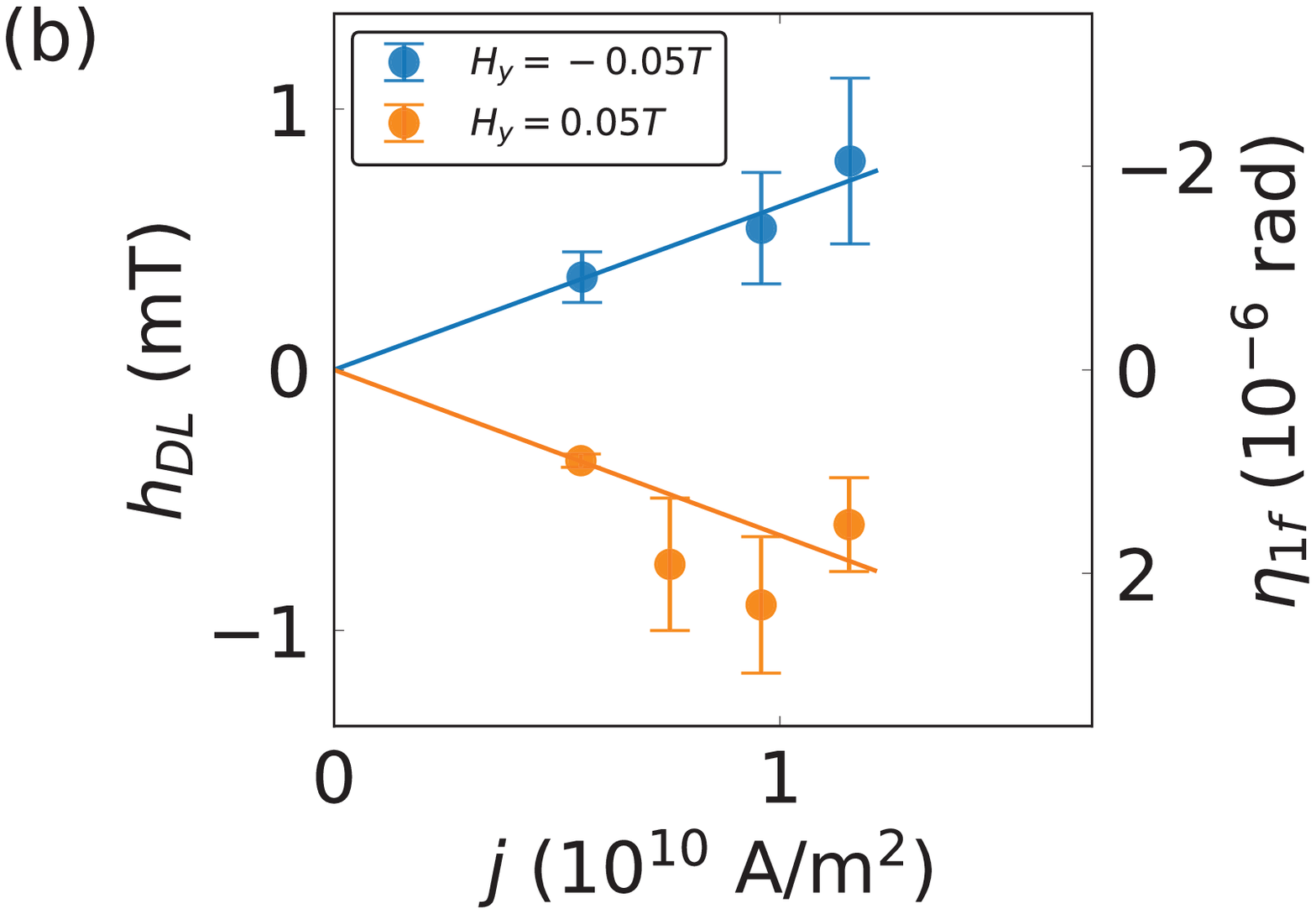}
		
	\includegraphics[width=0.49\linewidth]{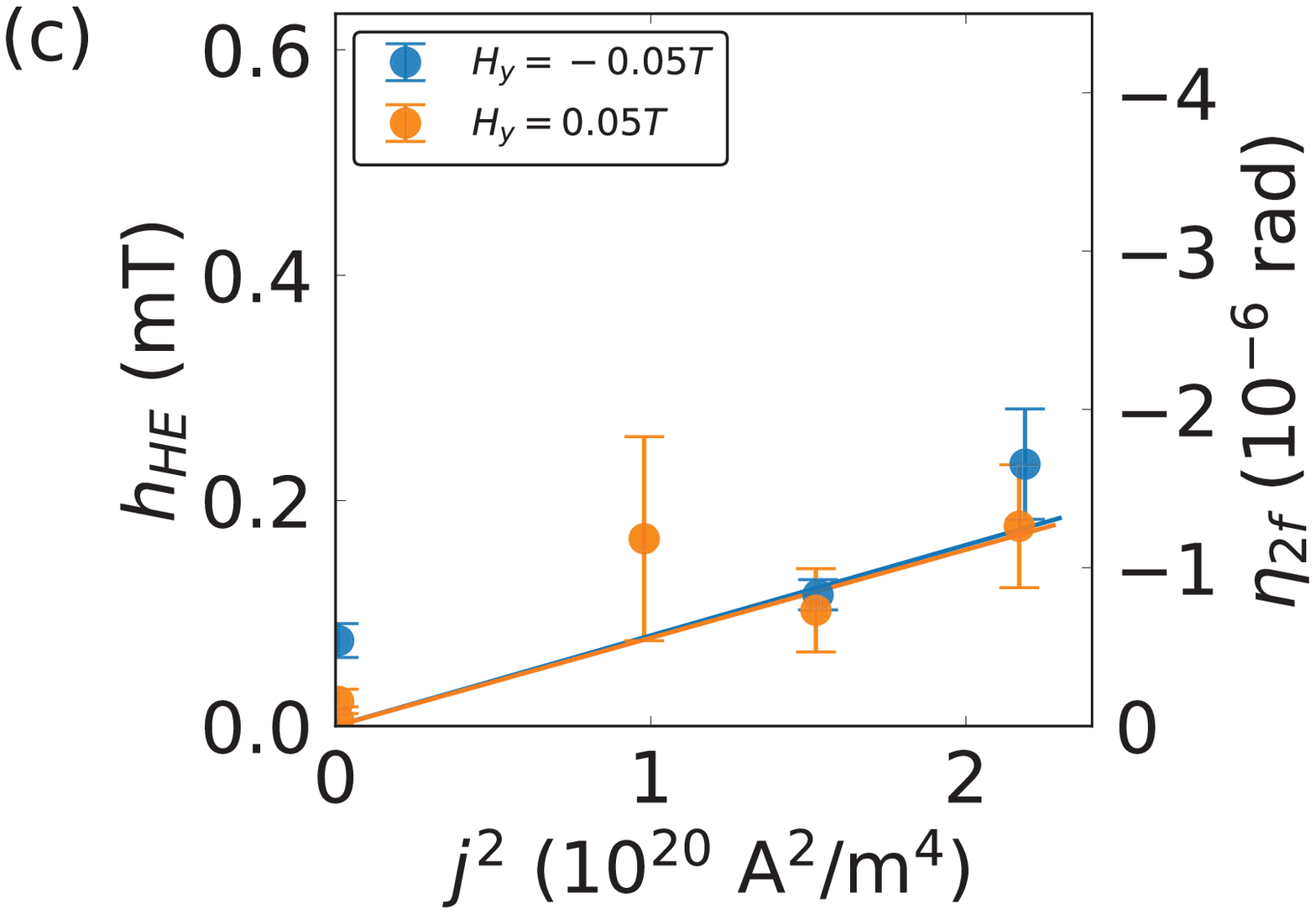}
	\includegraphics[width=0.49\linewidth]{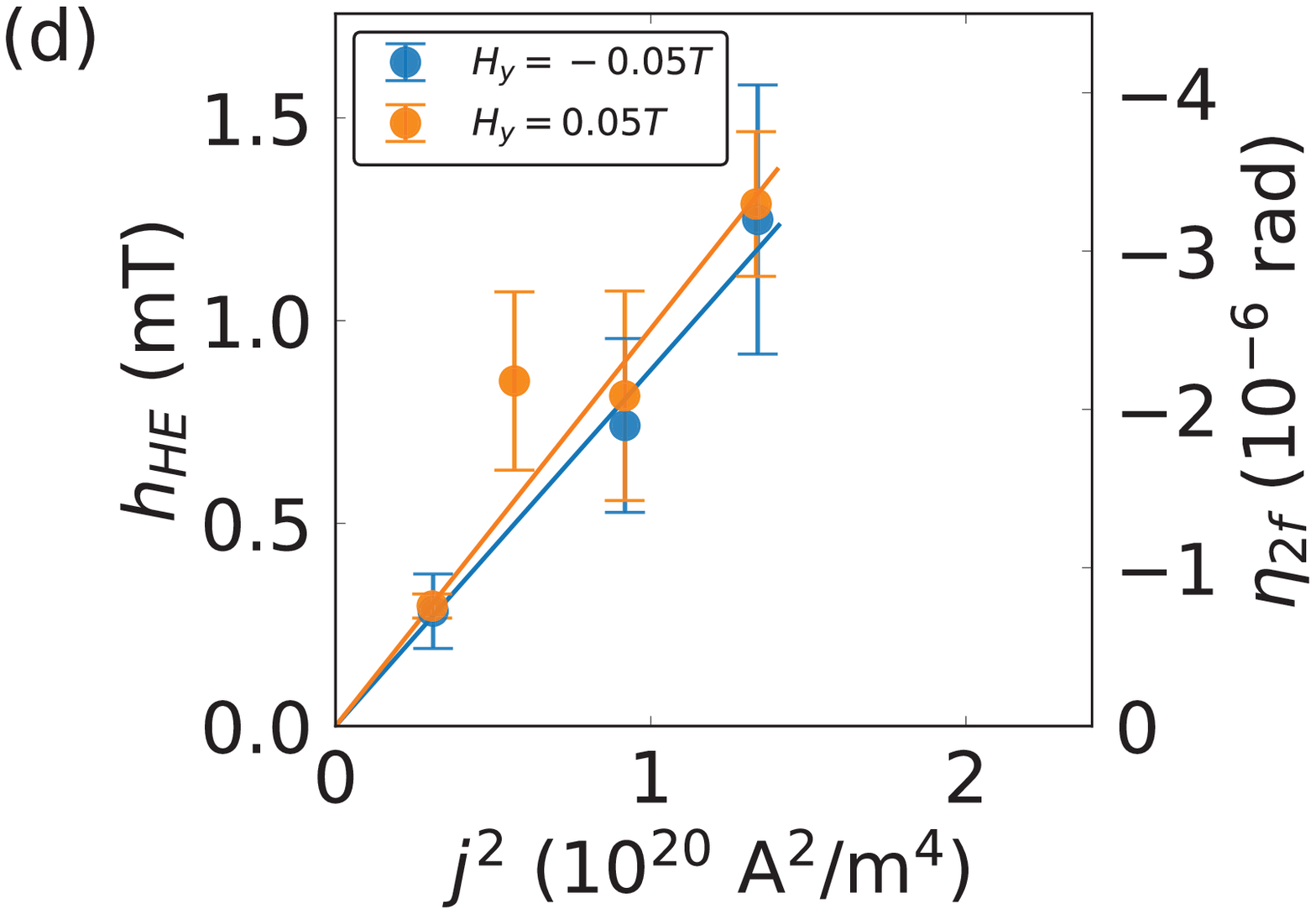}
	\caption{(a, b) $h_{DL}$ as a function of $j$ for samples A (a) and B (b). (c, d) $h_{HE}$ vs. $j^2$ for samples A (c) and B (d). $j$ is modulated with frequency $f$. Here $\tau =3$ s ($f=0.8\times10^{-2}$ Hz). The error bars are calculated from the covariance matrix of the fitting parameters. The blue and orange symbols represent effective fields when the magnetization of the FM layer points along $-y$ and $+y$, respectively, and the lines are linear fit of the data. 
	}
	\label{fig:terms}
\end{figure}

The effective fields $h_{DL}$ and $h_{HE}$ are shown in Fig. \ref{fig:terms} as a function of $j$ and $j^2$, respectively, for both samples A and B. As evident, $h_{DL}$ linearly scales with $j$ and changes its sign when the magnetization direction of the initial state is reversed. The sign of $h_{DL}$ is opposite for samples A and B, reflecting the opposite spin Hall angle of W and Pt. These features are consistent with the damping-like component of the current induced torque. The spin torque efficiency\cite{N2949} $\xi$ of samples A and B can be obtained from $h_{DL}$. Assuming a transparent interface, which provides the lower limit of the effective spin Hall angle, we use the following relation for the conversion\cite{N1896}:
\begin{equation}\label{key}
	\xi = \frac{(h_{DL}/j)\times M_{s} t}{\hbar/2e}
\end{equation}
where $M_s$ and $t$ are the saturation magnetization and thickness of the FM layer, respectively, $\hbar$ is the reduced Planck constant and $e$ is the electric charge. $\xi$ and $M_s$ for samples A and B are summarized in Table \ref{tab:efficiency}. For comparison, we have also measured the spin Hall magnetoresistance\cite{N2770,N2523,N2946,N2522} (SMR) of samples A and B using the Hall bars. 
These results are included in Table \ref{tab:efficiency}. We find that both techniques, magneto-optical detection ($\xi_{MOKE}$) and SMR ($\xi_{SMR}$), return $\xi$ that are comparable in magnitude. $\xi$ is consistently smaller when the former method is used: the reason behind this is unclear and requires further investigation.

\begin{table*}[t]
	\centering
	\caption{Spin torque efficiency $\xi$ obtained from the optical method described here ($\xi_{MOKE}$) and the SMR ($\xi_{SMR}$), saturation magnetization $M_s$ (from Ref. \cite{N3284}), resistivity of HM $\rho$ and conversion factor $\alpha$ that relates $\eta$ to effective field.\\}
	\begin{tabular}{crrrrr}
		\hline\hline  \multicolumn{1}{c}{$ \textrm{sample} $}  & \multicolumn{1}{c}{$ \xi_{MOKE} $} & \multicolumn{1}{c}{$ |\xi_{SMR}| $} & \multicolumn{1}{c}{$ M_{s} $(kA/m)} & \multicolumn{1}{c}{$ \rho(\mu\Omega $cm$) $} & \multicolumn{1}{c}{$ \alpha $(rad/T)} \\
		\hline A (Pt) & $0.135\pm0.007$ & $ 0.141\pm0.007 $                         & $1500$                  & $ 55 $                      & $(7.1\pm0.1)\times10^{-3}$                 \\
		B (W)         & -$0.19\pm0.01$ & $ 0.22\pm0.01 $                         & $1000$                  & $ 107 $                      & $ (2.56\pm0.04)\times10^{-3}$               \\
		\hline\hline    &                                                                                                                                                                                                  
	\end{tabular}
	\label{tab:efficiency}
\end{table*}

With regard to $h_{HE}$, we find that it scales with $j^2$ and its sign is the same regardless of the initial state of the magnetization direction and the HM layer used (i.e. samples A and B possess the same sign for $h_{HE}$). These results indicate that $h_{HE}$ is related to current induced (Joule) heating. The difference in the size of $h_{HE}$ for samples A and B may partly be explained by the difference in the film resistance. As noted in Table \ref{tab:efficiency}, the resistivity of the HM layer is nearly two times larger for sample B, giving rise to larger Joule heating (heating scales with the film resistance for a fixed current density; the geometry of the wire is the same for both samples). We note that, however, the absolute value of $h_{HE}$ depends on extrinsic factors, e.g. thermal contact between the sample and the sample stage, laser power and temperature of the environment. It is thus difficult to compare $h_{HE}$ between samples as these effects can vary. 

\begin{figure}[!t]
	\includegraphics[width=0.49\linewidth]{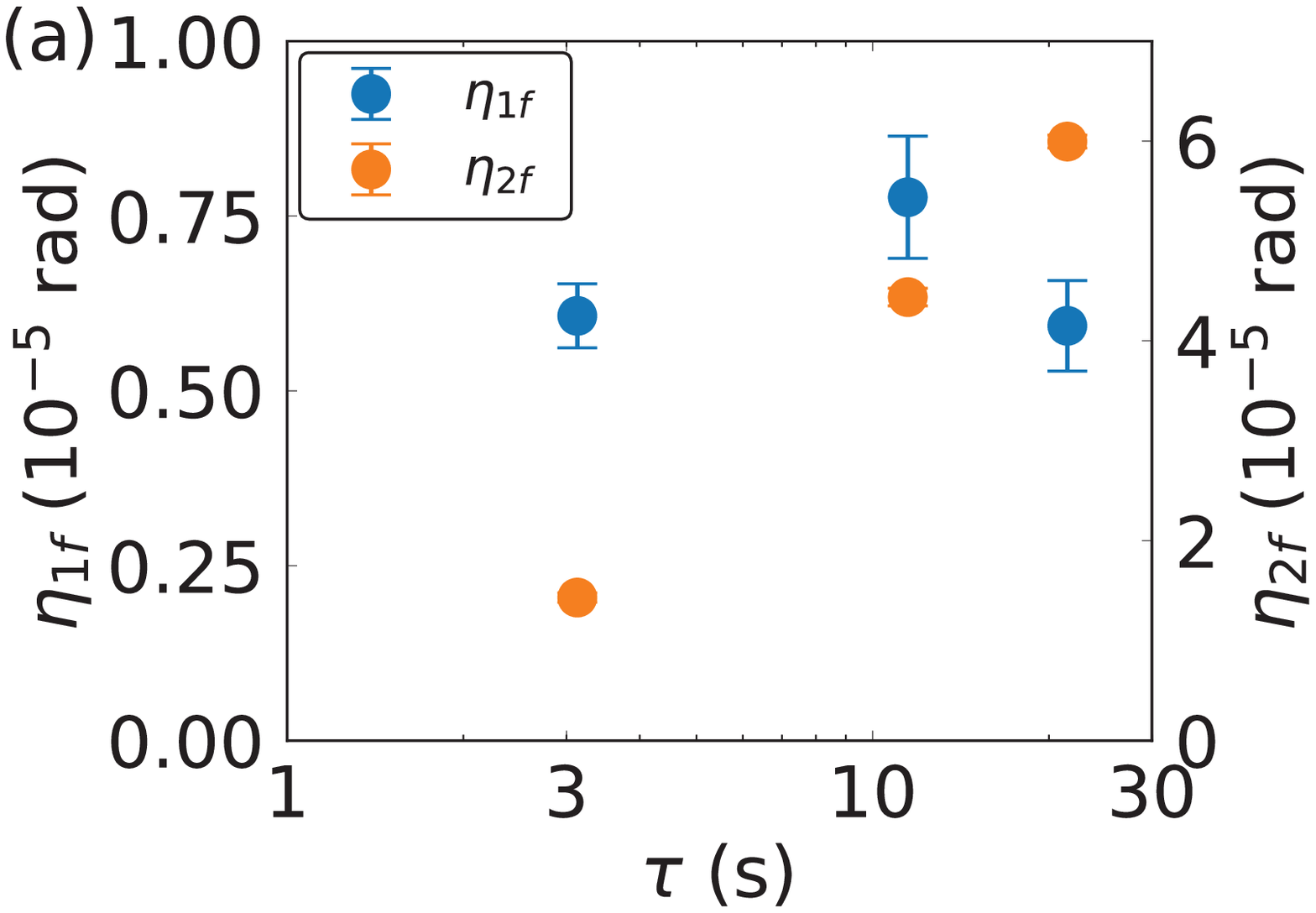}
	\includegraphics[width=0.49\linewidth]{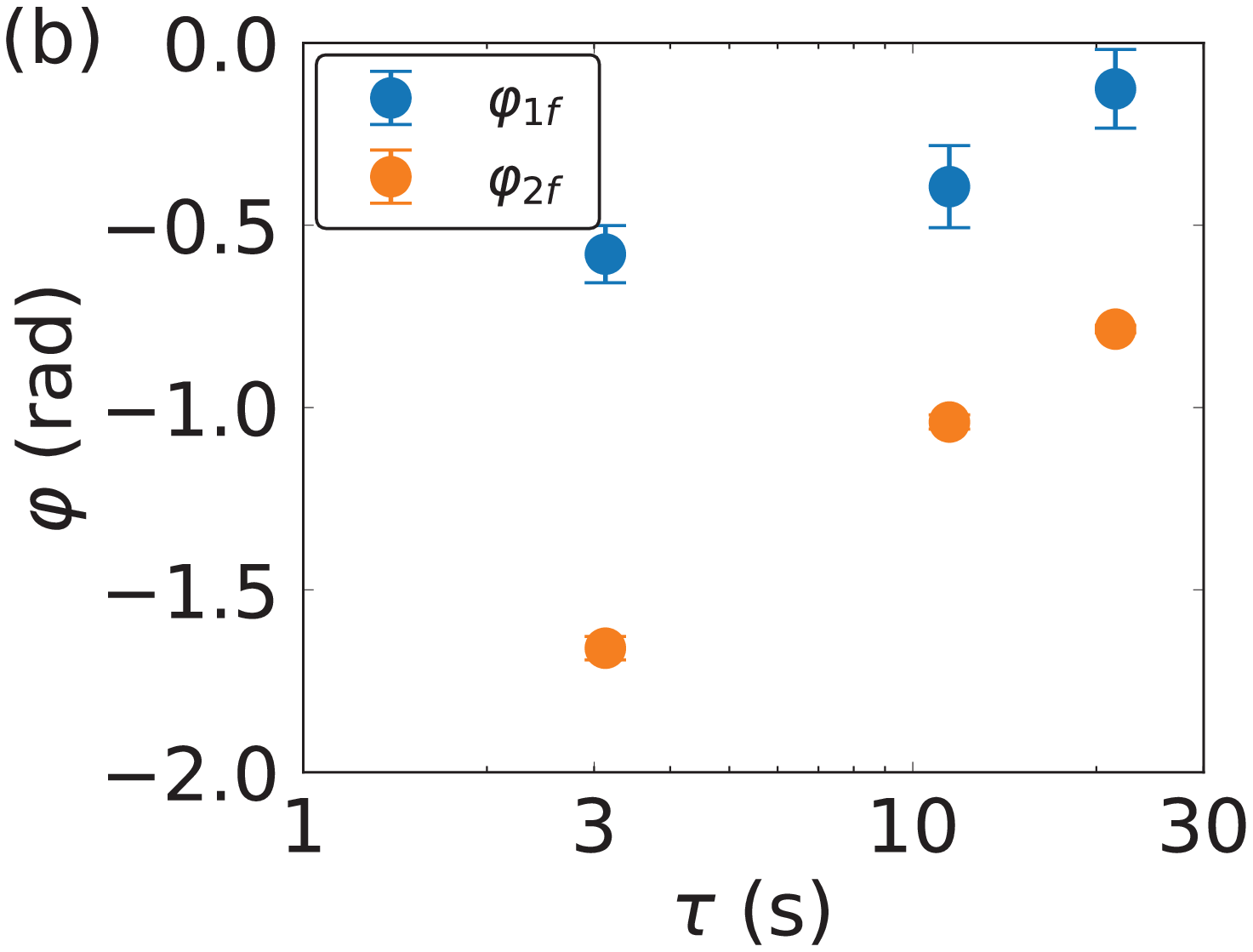}
	\caption{(a) Dependence of $\eta_{1f}$ and $\eta_{2f}$ as a function of $\tau$. (b) $\varphi_{1f}$ and $\varphi_{2f}$ vs. $\tau$. The blue and orange symbols represent components that are proportional to $f$ and $2f$, respectively. Results are from sample B and $ j \sim 10^{10}\mbox{A/m}^2 $. The error bars are calculated from the covariance matrix of the fitting parameters.}
	\label{fig:freq}
\end{figure}

We have also varied the duration of current application $\tau$ to study the degree of heating on the optical signal and estimation of the spin orbit torque. The $\tau$ dependence of $\eta_{1f}$ and $\eta_{2f}$ are shown in Fig. \ref{fig:freq}(a). Whereas $\eta_{1f}$ shows little change on $\tau$, $\eta_{2f}$ significantly increases with increasing $\tau$. We also plot the phase delay ($\varphi_{1f}$ and $\varphi_{2f}$) as function of $\tau$ in Fig. \ref{fig:freq}(b). Non-zero phase delay for small $\tau$ indicates that the heating process does not saturate in one cycle (within $\tau$). We find a large change in $|\varphi_{2f}|$ with increasing $\tau$ while changes in $|\varphi_{1f}|$ is small. These results indicate that the optical signal related to measurements of spin orbit torque ($\eta_{1f}$) is not significantly influenced by the degree of heating, i.e. by $\tau$. 

We next discuss the origin of $h_{HE}$. The Joule heating induced Kerr signal change $\frac{\eta_{2f}}{\eta_0}$ is, for example, $\sim 2 \times 10^{-3}$ from the results shown in Fig. \ref{fig:wave}(b), red dashed line. 
We consider there are two possible sources of $h_{HE}$: temperature dependent changes of the film thickness, including the silicon oxide layer of the substrate, and the oxide/film refractive index. Note that the Kerr signal observed here are enhanced due to optical interference effect that takes place within the oxide layer of the substrate\cite{N3337}. Upon Joule heating, the each layer undergoes thermal expansion and changes its thickness. In addition, the refractive index of each layer changes due to the temperature change. We use the effective refractive index approach to estimate changes in the Kerr signal\cite{N3238,N2103}. With the thermal expansion coefficients of the silicon oxide\cite{N3379}, we find $\frac{\eta_{2f}}{\eta_0} \sim 10^{-6}$ (1/K). 
(Contribution of the thermal expansion of the film on the Kerr signal is even smaller.) 
In contrast, changes in the refractive index\cite{N3380,N3381} causes $\frac{\eta_{2f}}{\eta_0} \sim 5 \times 10^{-4}$ (1/K): contribution from each layer is nearly the same. As the latter contribution is larger by orders of magnitude, we consider $h_{HE}$ is caused by Joule heating induced changes of the refractive index of the oxide/film. For a current density of $\sim 10^{10}$ A/m$^2$, we estimate it will require the temperature to change by $\sim 4$ K to observe the changes in $\frac{\eta_{2f}}{\eta_0}$ found here.

Finally, it has been reported recently that spin accumulation due to the spin Hall effect in non-magnetic metals can be detected using magneto-optics\cite{N3027,N2908,N3363,N3362}. To reduce spurious optical signal from Joule heating, the current density passed along the film has to be reduced to $\sim 10^7$ A/m$^2$ or smaller. As a consequence, the signal resolution has to be better than $10^{-8}$ rad\cite{N3363}. Using the optical detection scheme developed here, we consider it is possible to  separate optical signals due to spin accumulation and the Joule heating effect, allowing larger current to be passed to the film. Assuming that the optical signal due to spin accumulation scales with the current density, one estimates Kerr signal of $10^{-5}$ rad at the current density used here ($\sim 10^{10}$ A/m$^2$). Since the Kerr resolution of the setup here is $\sim 10^{-6}$ rad (see Fig. \ref{fig:wave}(b)), we consider it is possible to detect the signal due to spin accumulation.

In summary, we have studied using magneto-optics current induced spin orbit torque in HM/FM bilayers (HM=Pt, W, FM=CoFeB). Using a double modulation technique, optical signals arising from the magnetic system are separated from those due to current induced (Joule) heating. Although we find significant contribution from Joule heating on the optical signal at current density of $\sim 1 \times 10^{10}$ A/m$^2$, the spin torque efficiency vary little with the current density. The obtained spin torque efficiency agrees with that estimated using spin transport measurements of the same sample. We find that the Joule heating induced optical signal originates from changes in the temperature dependent refractive index of the film/silicon oxide layer. As the detection scheme developed here can separate magnetic and heating related optical signals, we consider this technique can be applied to study spin accumulation in metals and interfaces. 

\begin{acknowledgments}
This work was partly supported by JSPS Grant-in-Aid for Scientific Research (16H03853), Specially Promoted Research (15H05702) and the Center of Spintronics Research Network of Japan. 
\end{acknowledgments}

\bibliography{ref_050218}

\end{document}